\documentclass[twocolumn,showpacs,preprintnumbers,amsmath,amssymb]{revtex4-2} 
\usepackage{graphicx}
\usepackage{dcolumn}
\usepackage{bm}
\usepackage{tabularx}
\usepackage{ulem} 
\newcolumntype{M}{>{\centering\arraybackslash}m{1.85cm}}
\usepackage[export]{adjustbox}
\usepackage{float}
\usepackage{braket}
\usepackage{lipsum}
\usepackage{amsmath}
\graphicspath{{figure/}}
\usepackage{xcolor}   

\usepackage{caption}
\usepackage{subcaption}

\usepackage{hyperref}
\hypersetup{
    colorlinks=true, 
    linktoc=all,     
    urlcolor  = cyan,
    citecolor = blue,
    linkcolor=blue,  
}

\makeatletter
\newcommand{\colorcaption}[2][]{%
  \begingroup%
  \renewcommand{\@caption@fignum@sep}{ (Color online). }%
  \caption[#1]{#2}%
  \endgroup%
}
\makeatother
\newcommand{\orcid}[1]{\href{https://orcid.org/#1}{\hskip2pt\includegraphics[width=9pt]{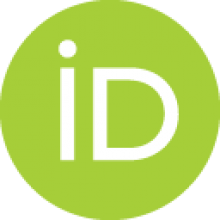}}}

\begin{document}

\title{ \textit{Ab initio} study of island of inversion in odd-$A$ nuclei: Structure of $^{31,33}$Mg }

\author{Subhrajit Sahoo\orcid{0000-0001-8000-2150}}
\email{s$\_$sahoo@ph.iitr.ac.in}
\affiliation{Department of Physics, Indian Institute of Technology Roorkee, Roorkee 247667, India}

\author{Praveen C. Srivastava\orcid{0000-0001-8719-1548}}
\email{ praveen.srivastava@ph.iitr.ac.in}
\affiliation{Department of Physics, Indian Institute of Technology Roorkee, Roorkee 247667, India}

\date{\hfill \today}

\begin{abstract}
We study the $N=20$ island of inversion region in the odd-$A$ Ne and Mg isotopes from the fundamental nuclear forces based on chiral two- and three-nucleon potentials. The state-of-the-art \textit{ab initio} valence space in medium similarity renormalization method was used for this purpose. Our study focuses on the evolution of single-particle states and discusses their transition into the island of inversion through particle-hole excitations across the $N=20$ shell gap. The computed low-lying states and magnetic moments are in good agreement with the experimental data. We presented the rotational band structures, established via $E2$ transitions, in $^{31}$Mg and $^{33}$Mg, which emerge from both normal and intruder configurations at low excitation energies. Our results suggest the presence of weak, moderate, and strongly prolate-deformed configurations at low energy in these  isotopes. The present work offers valuable insights into the configurations and shapes of low-lying states in nuclei within the island of inversion, enhancing our understanding of the structures of exotic nuclei from first principles.
\end{abstract}


\maketitle
\section{Introduction}
The shell gaps and deformation properties serve as key features for understanding the structure of complex atomic nuclei. Compared to their stable counterparts, these properties change remarkably in the isotopes lying far from the $\beta$-stability line of the nuclear chart \cite{SorlinReview}. The breakdown of the conventional $N=20$ magic number in neutron rich $Z=10-12$ isotopes constitutes a good example where the ground states are deformed, resulting from a diminished shell gap between $sd$- and $pf$- shell \cite{OtsukaReview}. This region is referred to as the ``island of inversion (IoI)" \cite{WarburtonIoI, Poves1994}, where multi particle-hole excitations across the narrowed shell gap dominate the ground-state configuration. The dissolution of the $N=20$ magic number and the emergence of collectivity with increasing neutron numbers in the IoI region have drawn significant interest from both experimental and theoretical studies \cite{SorlinReview, OtsukaReview, WarburtonIoI, Poves1994, BrownReview}. This region provides a rich testing ground for nuclear interactions, as well as theoretical models, to explain the structure of nuclei far from stability.

The evolution of single-particle states and their configurations are keys to understanding the transition into the island of inversion region. In contrast to even-even nuclei, the odd-mass systems display more complex structures, particularly near the $N=20$ shell gap due to the additional coupling between the unpaired nucleon with particle-hole excitations. In $^{31}$Mg, lying between the normal $^{30}$Mg and intruder $^{32}$Mg, shape coexistence is observed in which different shapes coexist within a narrow range of low-excitation energy \cite{31Mg_ShapeCo1, 31Mg_ShapeCo2}. The low-energy structure of $^{33}$Mg has garnered significant attention in the past years as the experimental measurements of its ground state spin-parity with different methods have produced conflicting results \cite{33Mg_gsSpin2001, 33Mg_gsSpin2002, 33Mg_gsSpin2008, 33Mg_gsSpin2010} and do not align with the measured ground state magnetic moments \cite{33Mg_gsMu2007, 33Mg_gsMu2011}. Later, the momentum distribution measurements from one-proton and one-neutron knockout reactions were found to be consistent with a $p$-wave shape and suggested a spin-parity assignment of $3/2^-$ for the ground state \cite{33Mg_gsSpin2021}. Although the spin-parities of excited states are still unclear or tentative,  both theoretical investigations and experimental studies have suggested shape coexistence \cite{ShapeCo_Mg33, ShapeCo_N21} and fingerprints of rotational band structures in $^{33}$Mg \cite{RotBand_Mg33}. While the even-even nuclei have been extensively studied (e.g., see Ref. \cite{OtsukaReview} for a recent review), the odd-mass isotopes in the IoI region are less explored. A systematic study of odd-mass isotopes approaching IoI is still limited. Theoretical studies on band structures and shape coexistence in nuclei within IoI region, based on underlying nuclear forces, are desirable for resolving experimental ambiguities and deepening our understanding of nuclear structure in exotic isotopes.

The phenomenological interactions—where one- and two-body matrix elements of the Hamiltonian are fitted with several experimental data—have shown varying degrees of success in describing odd-$A$ systems within the IoI \cite{SDPF-M, SDPF-U, SDPF-MU, SDPF-U-MIX}. However, despite significant progress in \textit{ab initio} methods, achieving a fully microscopic understanding of this region from the fundamental nuclear forces remained challenging due to the problem of intruder states associated with multi-valence-space Hamiltonians ($sd$- and $pf$-shell) \cite{RagnarReview}. To address this, the extended Kuo-Krenciglowa (EKK) method within Many-Body Perturbation Theory (MBPT) was employed to derive a multishell ($sd+pf$) valence space Hamiltonian (named as `EEdf1') from chiral two-nucleon (2N) and three-nucleon (3N) forces. However, it provides an accurate description of the IoI region only when the single-particle energies are adjusted empirically. Similarly, the recently developed angular momentum projected Coupled Cluster calculations with singles and doubles (CCSD) \cite{CC_projected} also struggle to reproduce the observed low-energy structure with the correct ordering of energy levels when applied to odd-$A$ systems \cite{IOI_CC} near the $N=20$ shell gap.

In recent years, the in-medium similarity renormalization group (IMSRG) \cite{RagnarReview, HeikoReview} has emerged as a powerful \textit{ab initio} many-body method for describing ground- and excited-state observables in open-shell nuclei. One of the recent works in IMSRG that combines the multi-reference IMSRG (MR-IMSRG) with the quantum-number projected generator coordinate method (PGCM), referred to as IM-GCM, has effectively captured the low-energy structure and shape coexistence in neutron-rich Mg isotopes \cite{IMSRG_GCM}. Another variant of IMSRG, the valence-space IMSRG (VS-IMSRG), non-perturbatively derives effective interactions for a chosen valence-space from chiral 2N and 3N forces \cite{IMSRG_RagnarPRC, IMSRG_RagnarPRL} and has shown promising results in describing a wide range of structure observables, such as binding energies, low-lying spectra, electromagnetic moments and transitions \cite{IMSRG_RagnarPRC, IMSRG_RagnarPRL, IMSRG_Miyagi, Miyagi2022PRC, MiyagiMomentsPRL}. The VS-IMSRG method has been extensively applied to explain the structural properties across light to heavier mass nuclei lying within a single \cite{Na_work_NPA, Li2023PLB, YuanN28PLB, YuanPRCL2024} as well as multiple major shells \cite{IMSRG_Miyagi, YuanN50PLB}.

In the present work, we have addressed the odd-$A$ isotopes of the island of inversion region within the VS-IMSRG framework. A brief overview of the VS-IMSRG approach for deriving multishell valence-space Hamiltonians is provided in Sec. \ref{Sec2}. The results are presented in Sec. \ref{Sec3}, where we have studied the evolution of the single-particle states approaching IoI and investigated the multi particle-hole configurations associated with them. Further, the structure and configuration of the odd-$A$ isotopes are probed through magnetic moments. Then we presented a detailed analysis of the low-energy structure of $^{31}$Mg and $^{33}$Mg that discusses their configurations and rotational band structures established via $E2$ transitions in both isotopes. The shapes of various states within low-excitation energy are also analyzed through quadrupole moments. Finally, the conclusions are summarized in Sec. \ref{Sec4}.

\section{Method} \label{Sec2}
Within the IMSRG framework \cite{RagnarReview, HeikoReview}, the many-body Schr\"{o}dinger equation is solved by applying continuous unitary transformations to the Hamiltonian. This is achieved via the SRG flow equation given by
\begin{equation} \label{eq:1}
\frac{dH(s)}{ds}=\left[ \eta(s),H(s)\right].
\end{equation}
Here, `$s$' is known as the flow parameter, and $\eta(s)$ is called the anti-hermitian generator, which performs the unitary transformations as $s$ approaches from $s=0$ to $s \rightarrow \infty$. To avoid computational complexities, the many-body Hamiltonian, derived from the chiral $NN$ and $NNN$ forces, is truncated at a two-body level, yielding the expression
\begin{equation} 
H=E_0(s)+\sum_{ab}f_{ab}(s){ \lbrace a^\dagger_a a_b \rbrace }+\frac{1}{4}\sum_{abcd}\Gamma_{abcd}(s){ \lbrace a^\dagger_a a^\dagger_b a_d a_c \rbrace },
\end{equation}
where $E_0$, $f_{ab}$, and $\Gamma_{abcd}$ are zero, one, and two-body terms, respectively, that capture the contribution of $NNN$ forces effectively through ensemble normal ordering \cite{IMSRG_RagnarPRL}. This is known as IMSRG(2) approximation. In practice, Eq. \eqref{eq:1} is solved using the Magnus formulation of IMSRG \cite{IMSRG_Magnus} to decouple multi-valence-space Hamiltonians employing the generator from Ref. \cite{IMSRG_Miyagi}.  It can be noted that this generator differs from those commonly used in the IMSRG method \cite{RagnarReview, HeikoReview} only by introducing an additional energy shift $\Delta$ in the denominator, which helps to solve issues associated with the decoupling of multishell Hamiltonians (for details, see Ref. \cite{IMSRG_Miyagi}). Typically, the size of $\Delta$ can be approximated as $\approx 41A^{-1/3}$ or of the order of 10 MeV for the present case, and varying $\Delta$ from 10 to 20 MeV produces minor differences in the calculated energies. The present calculations adopt $\Delta=$10 MeV.
To eliminate spurious states or center-of-mass (c.m.) contamination arising in multi-valence-space calculations, the Gl\"{o}ckner-Lawson term `$\beta H_{c.m}$' is added to the intrinsic Hamiltonian $H_{intr}$ at the beginning. Then, the Hamiltonian $H=H_{intr}+\beta H_{c.m}$ is consistently evolved through the SRG flow equation within IMSRG(2) approximation \cite{IMSRG_Miyagi}.

\begin{figure}
    \centering
    \includegraphics[scale=0.65]{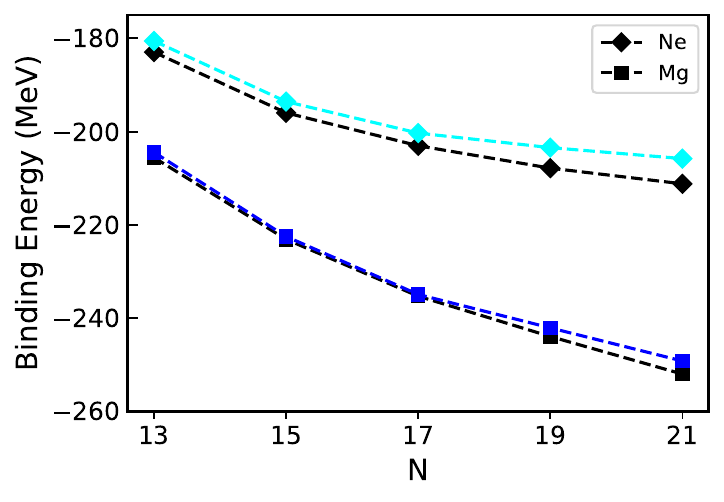}
    \caption{Experimental \cite{NNDC_Sn} (black) and calculated (colored) binding energies in odd-$A$ Ne and Mg isotopes.}
    \label{fig:gs_energy}
\end{figure}

\begin{figure*}
\includegraphics[scale=0.70]{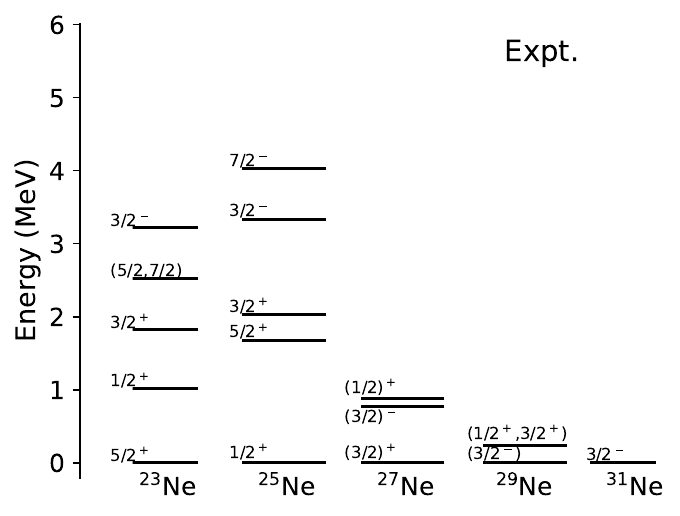} 
\hspace{3mm}
\includegraphics[scale=0.70]{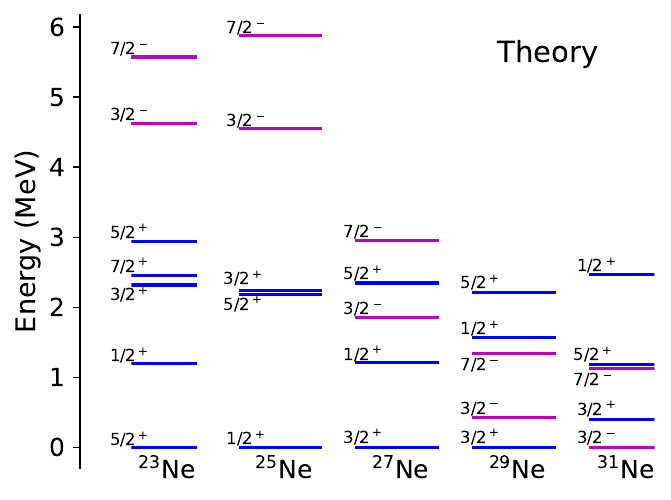}
\caption{Experimental \cite{NNDC, 29Ne_Liu2017} and calculated energy levels in Ne isotopes.}
\label{fig:statesNe}
\end{figure*}
\begin{figure*}
\includegraphics[scale=0.70]{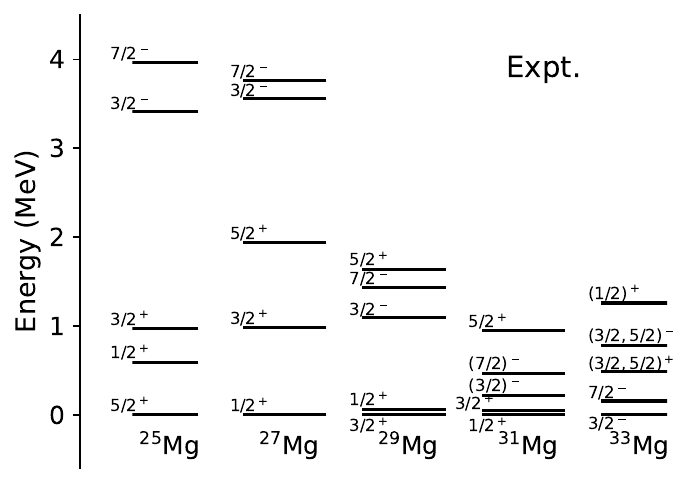} 
\hspace{3mm}
\includegraphics[scale=0.70]{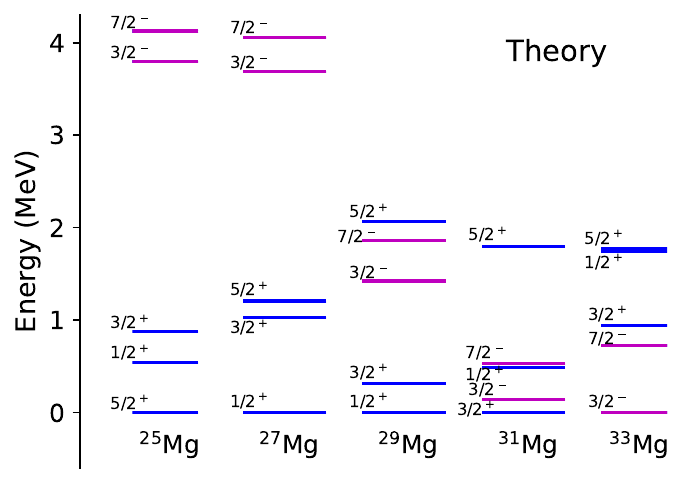}
\caption{Experimental \cite{NNDC, 29Mg_Matta2019} and calculated energy levels in Mg isotopes.}
\label{fig:statesMg}
\end{figure*}

In the present work, we have employed the chiral $NN$ and $NNN$ potentials labeled by EM 1.8/2.0 interaction \cite{EMinteraction1, EMinteraction2}, which remarkably reproduces the binding energies and spectroscopy of nuclei up to mass $A \approx 200$ \cite{Miyagi2022PRC}. The $NN$ component of this interaction is derived from a next-to-next-to-next-to-leading order (N$^3$LO) 2N potential \cite{EM2NPotential}, softened via SRG evolution to a momentum resolution scale of $\lambda = 1.8$ fm$^{-1}$, while the $NNN$ component corresponds to next-to-next-to-leading order (N$^2$LO) 3N potential with a momentum cutoff of $\Lambda = 2.0$ fm$^{-1}$. The VS-IMSRG calculations are carried out in a single particle basis with $e_{max}=12$ of frequency $\hbar \omega=16$ MeV, and 3N matrix elements are truncated at $E_{3max} = 24$ following the storage scheme described in Ref. \cite{Miyagi2022PRC}, ensuring convergence. Subsequently, the effective Hamiltonians are decoupled for a valence space for protons spanning $sd$-shell and neutrons in $sd$-shell plus $f_{7/2}$ and $p_{3/2}$ orbitals above the $^{16}$O core. The VS-IMSRG code of Ref. \cite{IMSRGCodeRagnar} is used for this purpose. The corresponding $E2$ and $M1$ operators are also consistently transformed within VS-IMSRG, eliminating the need for effective charges or $g$-factors. Finally, the Hamiltonians are diagonalized, and transition densities are extracted using the K-SHELL code \cite{KSHELL}. We observed that the Hamiltonians show minor differences by varying $\beta$ (for $\beta=2$, 3, and $4$) or remain nearly independent of $\beta$ for the nuclei under consideration, implying that they are devoid of c.m. contamination. In the following sections, we reported the VS-IMSRG results corresponding to $\beta=3$.

\begin{figure*}
\includegraphics[scale=0.58]{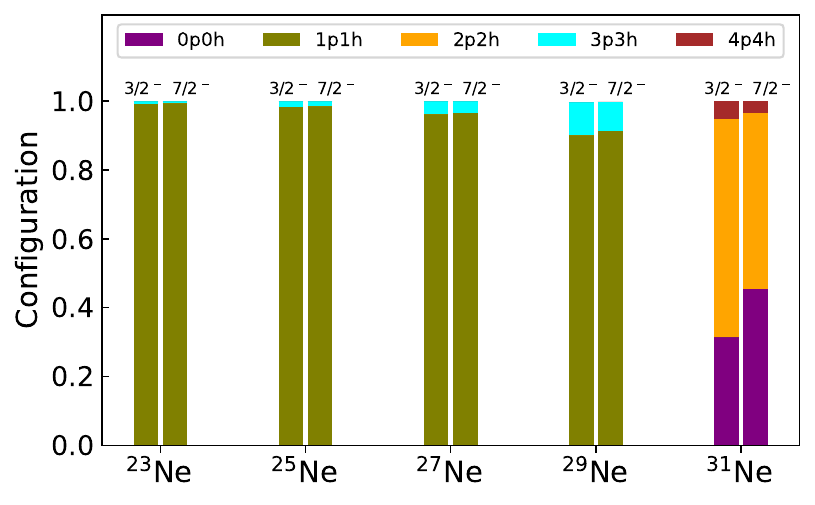} 
\hspace{2mm}
\includegraphics[scale=0.58]{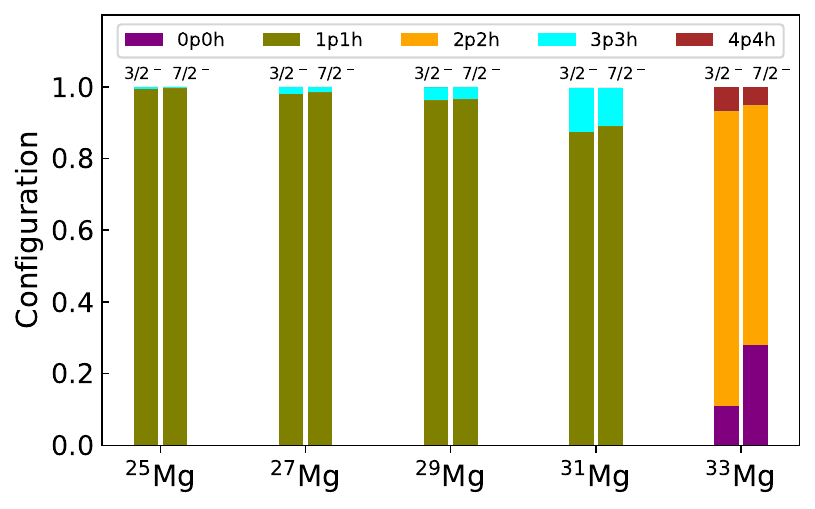}
\caption{Multi particle-hole configurations of the negative parity states in odd-$A$ Ne and Mg isotopes.}
\label{fig:phExcitationsNegNeMg}
\end{figure*}

\begin{figure*}
\includegraphics[scale=0.58]{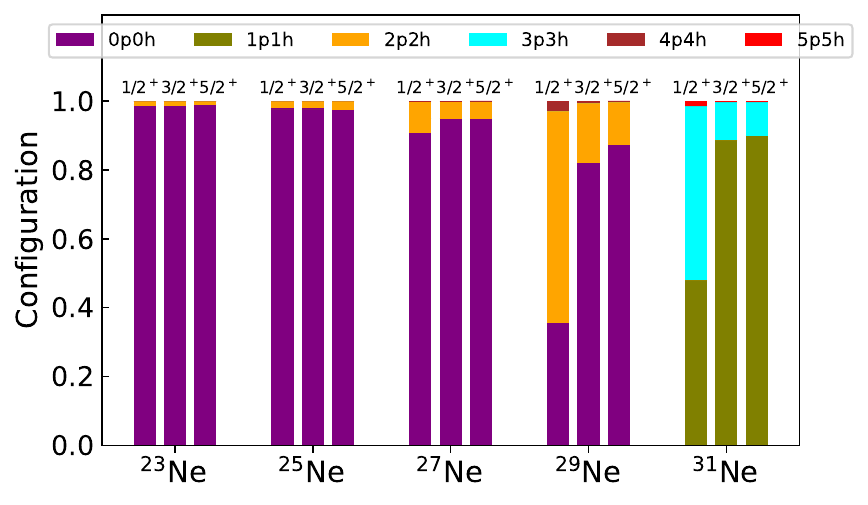} 
\hspace{2mm}
\includegraphics[scale=0.58]{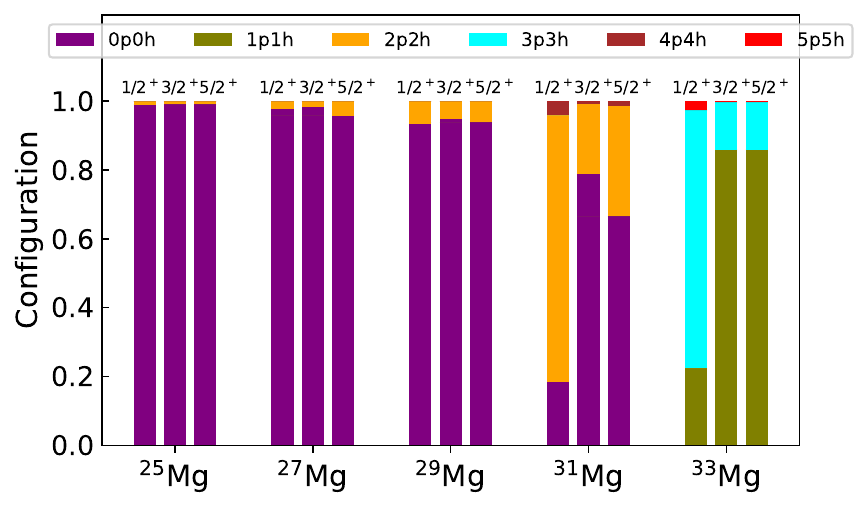}
\caption{Multi particle-hole configurations of the positive parity states in odd-$A$ Ne and Mg isotopes.}
\label{fig:phExcitationsPosNeMg}
\end{figure*}

\section{Results and Discussion} \label{Sec3}
First, we have shown the ground state binding energies of odd-$A$ Ne and Mg isotopes in Fig. \ref{fig:gs_energy}. They are in excellent agreement with the experimental data. Then, the spectroscopy of low-lying states in the Ne and Mg chains, obtained from VS-IMSRG calculations, are illustrated through Fig. \ref{fig:statesNe} and Fig. \ref{fig:statesMg}, respectively. Both figures systematically compared the theoretical results with their corresponding experimental counterparts. The calculated excited states are found at higher energies relative to the experimental data, likely due to the IMSRG(2) truncation employed. Recent studies incorporating three-body corrections into IMSRG(2) using factorized approximations \cite{IMSRG3f2} and IMSRG(3)-$N^7$ approximation \cite{IMSRG3N7_Heinz2021, IMSRG3N7_Ragnar, IMSRG3N7_Heinz2024}, at a higher computational cost, have shown a systematic reduction in the excitation energies.\\

\noindent
\textbf{Ne isotopes:} In Fig. \ref{fig:statesNe}, the experimental data of $^{23}$Ne indicate an energy level with tentative spins of $5/2$ and $7/2$ situated between the $3/2^+$ and $3/2^-$ states. Our calculated results suggest two distinct states, $5/2^+$ and $7/2^+$, within this range, with the $7/2^+$ state being lower in energy. The order of the single-particle states in Ne isotopes is well reproduced from VS-IMSRG calculations. However, only in the case of $^{27}$Ne, the $1/2^+$ and $3/2^-$ states appear inverted compared to the experimental prediction. The extracted momentum distribution suggests that the ground-state spin of $^{29}$Ne could be either $3/2^+$ or $3/2^-$ \cite{29Ne_Liu2017}. While $\beta$-decay measurements indicate a ground-state spin of $3/2^+$ \cite{29Ne_gsbetaDecay}, $1n$ removal studies strongly support $3/2^-$ as the ground-state spin in $^{29}$Ne \cite{29Ne_gs1nRemoval}. The VS-IMSRG calculations predict $3/2^+$ as the ground state of $^{29}$Ne with a low-lying $3/2^-$ as the first excited state. Additionally, we have reported several low-lying excited states predicted by VS-IMSRG in $^{29}$Ne and $^{31}$Ne, for which experimental data are currently unavailable.

\noindent
\textbf{Mg isotopes:} The yrast positive and negative parity states in $^{25}$Mg and $^{27}$Mg, obtained from VS-IMSRG calculations, show good agreement with the experimental data. However, in $^{29}$Mg, the calculated ground and first excited states appear inverted compared to the experiment. In $^{31}$Mg, the experimentally determined first excited state, $3/2^+$ at 50 keV, emerges as the ground state in the VS-IMSRG results, and the true ground state $1/2^+$ is found at 486 keV. This suggests that IMSRG calculations beyond IMSRG(2) truncation \cite{IMSRG3N7_Heinz2024, IMSRG3_52K} may be necessary to achieve better agreement with experimental observations. In contrast, the spin-parity of the calculated ground state, as well as the first excited state, closely aligns with the experimental measurements in $^{33}$Mg.

The appearance of $3/2^-$ below $7/2^-$ state indicates that the $\nu p_{3/2}$ orbital is lower in energy than the $\nu f_{7/2}$ in both Ne and Mg isotopic chains. From Fig. \ref{fig:statesNe} and Fig. \ref{fig:statesMg}, we can see a considerable drop in the energies of $3/2^-$ and $7/2^-$ states ($E(3/2^-)$ and $E(7/2^-)$) as we approach the $N=20$ shell gap. Ultimately, the $3/2^-$ state becomes the ground state in $^{31}$Ne and $^{33}$Mg. A steep change in the $E(3/2^-)$ and $E(7/2^-)$ can be observed at $N=17$ and 19 in both Ne and Mg chains, signaling a reduction in the energy gap between $sd$- and $pf$-shell and the onset of the island of inversion region.

\noindent
\textit{Particle-hole excitations:} Then, we study the multi particle-hole ($mpmh$) excitation probabilities associated with the yrast states through Fig. \ref{fig:phExcitationsNegNeMg} and Fig. \ref{fig:phExcitationsPosNeMg}, where $m$ denotes the number of particles excited across the $N=20$ shell gap. The probabilities of these $mpmh$ configurations are obtained by diagonalizing the multishell Hamiltonians within the full model space considered here.
Fig. \ref{fig:phExcitationsNegNeMg} shows that both $3/2^-$ and $7/2^-$ are primarily characterized by normal ($1p1h$) configurations, with a growing contribution from the intruder or $3p3h$ configurations as the $N=20$ shell gap is approached. Although the probability of the $3p3h$ configurations is small, it approximately doubles in the successive odd isotopes from $N=13$ to $N=19$. At $^{31}$Ne and $^{33}$Mg, where the $N=20$ shell gap collapses, the average no. of neutrons promoted from $sd$ to $pf$- shell is $\sim 1.5-1.9$ and $\sim 1.2-1.6$ in the $3/2^-$ and $7/2^-$ states, respectively. Both of these states are dominated by the intruder $2p2h$ configurations.

\begin{figure*}
\includegraphics[scale=0.74]{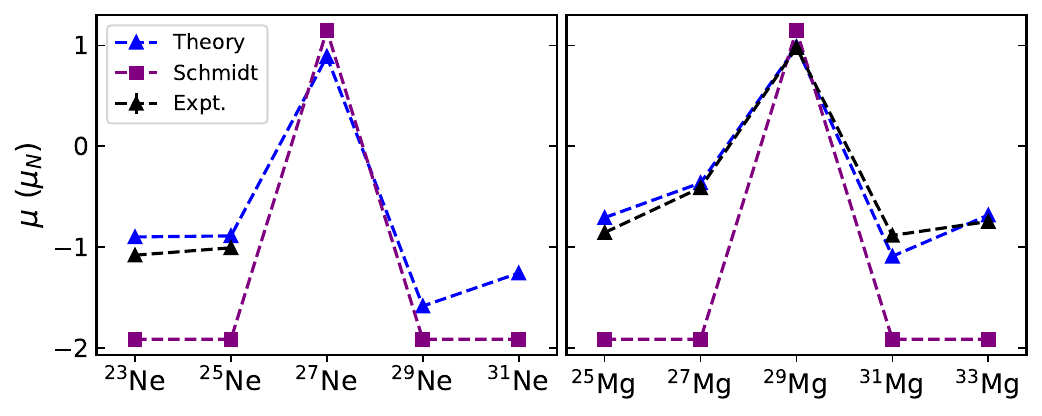} 
\caption{ Magnetic moments of the ground states in odd-$A$ Ne and Mg isotopes.}
\label{fig:GS_momentsNeMg}
\end{figure*}

\begin{figure*}
\includegraphics[scale=0.72]{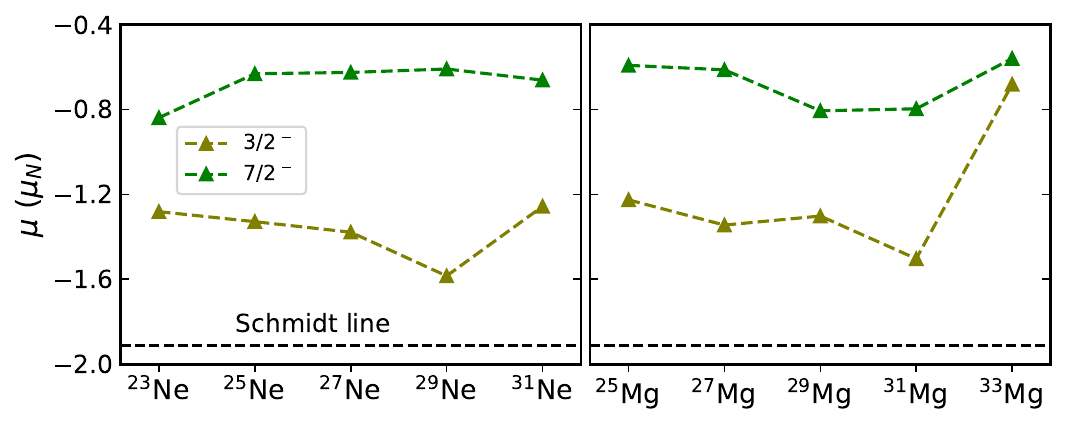} 
\caption{Magnetic moments of the negative parity states in odd-$A$ Ne and Mg isotopes.}
\label{fig:NegStates_momentsNeMg}
\end{figure*}

As observed in Fig. \ref{fig:phExcitationsPosNeMg}, the positive parity states exhibit a behavior similar to that of the negative parity states from $N=13$ to 17 and are characterized by normal configurations. At $N=19$, the $3/2^+$ and $5/2^+$ states have relatively larger intruder configurations compared to those in the negative parity states. However, unlike the negative parity states, these states are found to have dominant normal ($1p1h$) configurations at $N=21$. But, the $1/2^+$ state is observed to be predominantly of intruder character at both $N=19$ and 21. We see that, while the $3/2^+$ and $5/2^+$ states are still dominated by normal configurations, the $1/2^+$ state becomes completely of intruder nature in the island of inversion region.

Thus, within the low-excitation energy range, both positive and negative parity states exhibit different levels of intruder configurations around $N=20$. The probabilities of higher particle-hole excitations, such as $4p4h$ or $5p5h$, are found to be negligible. In contrast, they are significant above $N \geq 18$ in the EEdf1 interaction, where the two-body matrix elements (tbmes) are derived from the same chiral forces \cite{EEdf1}. Interestingly, the ground state ($3/2^-$) of $^{33}$Mg is found to be primarily of $2p2h$ nature across the $N=20$ shell gap when the same chiral potentials are employed in the IM-GCM approach.
This is consistent with the present results. Therefore, the discrepancies in particle-hole excitation probabilities between VS-IMSRG and EEdf1 results can be attributed to the fitting of single-particle energies made in the EEdf1 interaction. 
This adjustment likely alters the shell gap size and enhances the $ph$ excitations across it.  

\begin{figure*}[htbp]
    \raisebox{-0.5cm}{\includegraphics[width=0.44\textwidth]{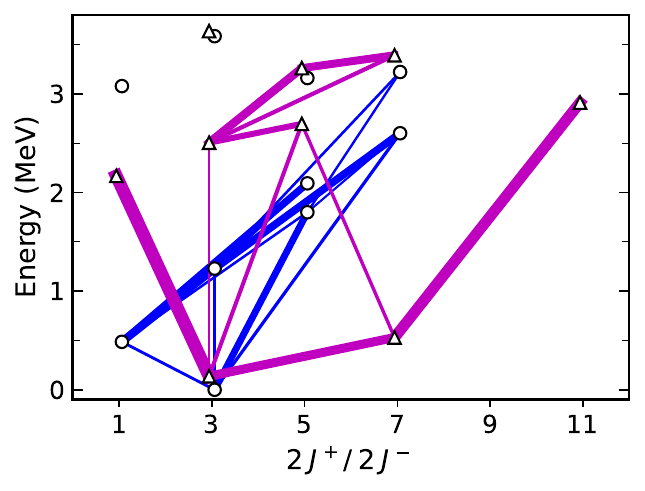}}
    \hspace{0.02\textwidth}  
    \raisebox{0.2cm}{\includegraphics[width=0.43\textwidth]{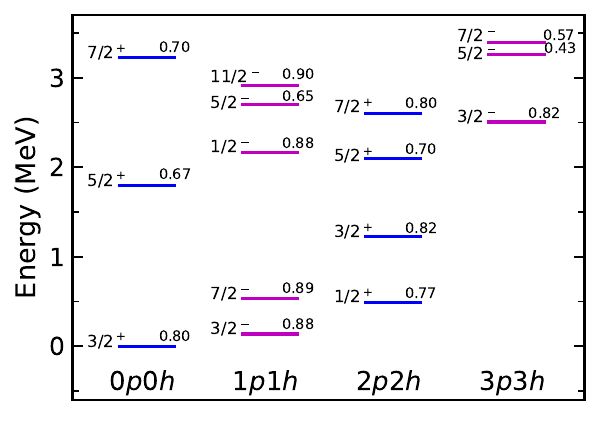}}
    \caption{ Left panel:- $E2$ map of $^{31}$Mg where excitation energies of positive (negative) parity states are plotted as circles (triangles) against their spins. The $E2$ transition strengths are shown through solid lines (above a threshold value of $8e^2fm^4$), with line widths proportional to $B(E2)$ strengths. Right panel:- Excited states of $^{31}$Mg, where each state is labeled with its spin on the left and particle-hole ($ph$) excitation probability on the right.}
    \label{fig:31MgE2mapAndStates}
\end{figure*}

\textit{Magnetic moments}: The electromagnetic observables, such as magnetic moments ($\mu$), serve as sensitive parameters to probe the intrinsic structure and configuration of odd mass isotopes. In Fig. \ref{fig:GS_momentsNeMg}, we present the ground state magnetic moments in odd-$A$ Ne and Mg isotopes. Here, by ground states, we refer to those observed experimentally and their theoretical counterparts. These values are also compared with the corresponding Schmidt values, which represent the single-particle magnetic moments. The ground state moments at $N=13$ and 15 deviate from the Schmidt values by $\sim 1.2$ units, implying large configuration mixing in them. But the ground state ($3/2^+$) magnetic moments at $N=17$ are much closer to the Schmidt line, with a difference of only about 0.2 units. This indicates that the $3/2^+$ state has a single-particle nature and predominantly appears from $[0^+ \otimes \nu (d^1_{3/2})]$ configuration in $^{27}$Ne and $^{29}$Mg. Further notable deviations are observed between the calculated and Schmidt values at $N=19$ and 21 in Mg, while these differences are comparatively smaller in Ne. This suggests that within the island of inversion, the ground-state wave functions in Mg are more mixed compared to those in Ne. Within VS-IMSRG calculations, all the observed ground state magnetic moments, wherever available, and their systematic trends are well reproduced.

The evolution and configurations of the $3/2^-$ and $7/2^-$ states in the odd-mass nuclei are key features of the island of inversion region. We investigate the nature of these states through their magnetic moments in both Ne and Mg isotopes. The calculated magnetic moments of the $3/2^-$ and $7/2^-$ states are illustrated in Fig. \ref{fig:NegStates_momentsNeMg}. Their values suggest that they are far away from the single-particle picture, particularly the $7/2^-$ state. The magnetic moments of the $3/2^-$ state differ from the single-particle value by $\sim 0.5-1.0$. This difference grows to $\sim 1.0-1.5$ in the $7/2^-$ states, and they appear from larger configuration mixing. The magnetic moments of the $7/2^-$ states remain almost constant across the Ne isotopic chain. However, a modest decrease is observed in the magnetic moments of the $3/2^-$ and $7/2^-$ states in Mg, as well as in the $3/2^-$ states of Ne, as we approach the $N=20$ shell gap. This indicates that with increasing $N$, certain configurations corresponding to filled $s$- or $d$- orbitals become dominant, which leads to an overall reduction in the configuration mixing towards $N=19$. We further noticed a sharp transition in the magnetic moments from $N=19$ to 21 or across the $N=20$ shell gap.  It is important to note that the present calculations of magnetic dipole moments do not include the effects of two-body currents (2BC), which have been shown to have a non-negligible contribution on the magnetic moments of near doubly magic nuclei \cite{2BC_Acharya, 2BC_Miyagi}. Investigating the impact of 2BC on magnetic moments in the IoI region will be an interesting aspect of future works.

Thus, we see that, within low-excitation energy, both positive and negative parity states coexist around $N=20$. The multi-particle hole excitations associated with these states and their configurations have markedly changed from $N=19$ to 21 or across the shell gap, indicating possible shape coexistence or shape transition. Around $N=20$, the Ne isotopes are more exotic; for instance, the neutron separation energy ($S_n$) in $^{31}$Ne is as low as 170 keV \cite{NNDC_Sn}. In such cases, continuum effects are expected to play a dominant role, and shell model calculations without the continuum effects may not be reliable. But, the excited states in $^{31,33}$Mg are well bound, and the neutron separation energy lies around $\sim 2.3$ MeV. Studying the band structures and their underlying configuration in these isotopes can provide intriguing insights into the nuclear structure within the island of inversion. Here, we discuss the structure of $^{31,33}$Mg in detail.
\vspace{1mm}

\begin{figure*}[htbp]
    \raisebox{-0.5cm}{\includegraphics[width=0.44\textwidth]{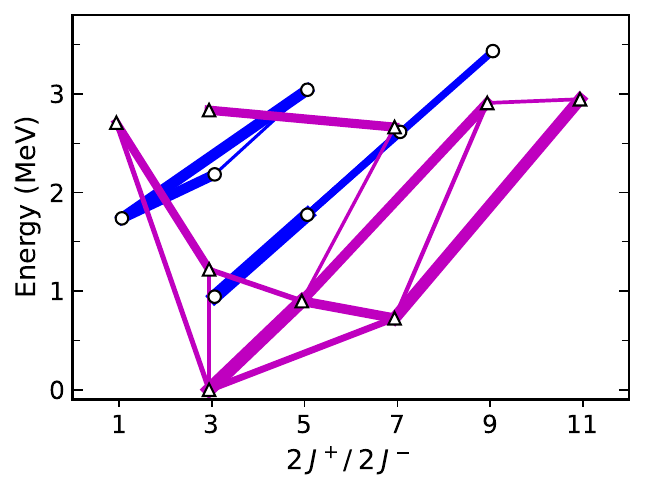}}
    \hspace{0.02\textwidth}  
    \raisebox{0.2cm}{\includegraphics[width=0.43\textwidth]{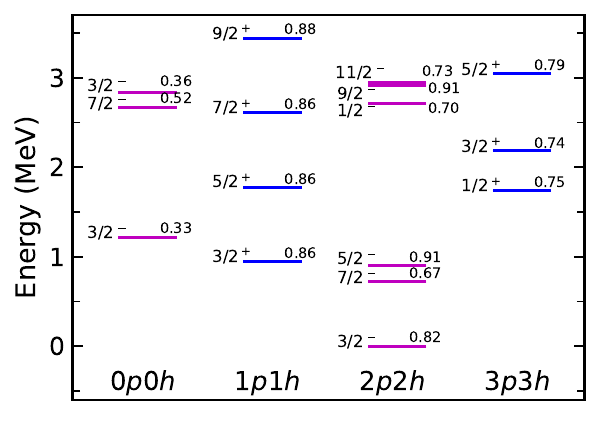}}
    \caption{ Same as Fig. \ref{fig:31MgE2mapAndStates} but for $^{33}$Mg. }
    \label{fig:33MgE2mapAndStates}
\end{figure*}

\noindent
\textbf{$^{31}$Mg:} The low-lying excited states of $^{31}$Mg, appearing from different $ph$ excitations are shown on the right panel of Fig. \ref{fig:31MgE2mapAndStates}. Each state is labeled with its spin-parity and the corresponding $ph$ excitation probability. The left panel of Fig. \ref{fig:31MgE2mapAndStates} depicts the $E2$ map between these states. The $E2$ maps are particularly useful for identifying the band structures within nuclei. In general, the states within a band are connected by stronger $E2$ transitions compared to those between bands. The VS-IMSRG calculations with evolved $E2$ operators underestimate the $E2$ strengths as well as quadrupole moments, primarily due to the lack of many-particle, many-hole excitations beyond the model space, as reported in Refs. \cite{E2_strength_22Mg, E2_strength_vsimsrg}. However, they provide a reasonable description of shell structure by reproducing the observed $B(E2)$ trends, and the relative $E2$ strengths are always meaningful \cite{IMSRG_Miyagi, YuanPRCL2024, YuanN50PLB}. Fig. \ref{fig:31MgE2mapAndStates} illustrates that states originating from similar $ph$ configurations constitute a band.
The $^{31}$Mg exhibits diverse band structures within low-excitation energy. The ground-state band connects the $3/2_1^+$, $5/2_1^+$, and $7/2_2^+$ states of normal configurations. While the $E2$ strength for the transition from $3/2_1^+$ to $5/2_1^+$ is considerable, it is relatively small $\sim 6–12$ $e^2 \mathrm{fm}^4$ for the transitions $3/2_1^+ \rightarrow 7/2_2^+$ and $5/2_1^+ \rightarrow 7/2_2^+$. A negative parity band, characterized by robust $E2$ transitions and connecting the $1p1h$ states, and a positive parity band, comprising $2p2h$ states, also coexist with the ground state band. This positive parity band with its sequences of states $1/2_1^+$, $3/2_2^+$, $5/2_2^+$ and $7/2_1^+$ are connected with enhanced $E2$ transitions except the $7/2_1^+ \rightarrow 5/2_2^+$ transition, which is hindered. Notably, the $7/2_1^+$ state directly decays to $3/2_2^+$ instead of $5/2_2^+$. Further, the $3/2_2^-$ in $^{31}$Mg serves as the band head of another band linked with $5/2_2^-$ and $7/2_2^-$ states and is distinguished by strong $E2$ transitions.

Recently, the band structures of odd-$A$ nuclei in the island of inversion region have been addressed from \textit{ab inito} CCSD calculations using angular momentum projection guided by Nilsson's model \cite{IOI_CC} and MR-IMSRG integrated with quantum-number projected generator coordinate method \cite{IMSRG_GCM}. In these studies, the computed wave functions have a well-defined $K^{\pi}$ value, $K$ being the angular momentum projection, and $\pi$ denotes the parity. However, the $K$ number can not be directly obtained from the valence-space calculations, but the $B(E2)$ strengths between the states of a rotational band can guide us to the $K$ value associated with that band. 
 For instance, if the $3/2_1^-$, $7/2_1^-$, and $5/2_1^-$ states belong to a rotational band with a fixed $K$ value, then their $B(E2)$ ratios can be expressed in terms of Clebsch-Gordan (C.G.) coefficients as \cite{KnumPRC2015, KnumNPA2018, KnumPRC2024} 
\begin{equation}
    R \equiv \frac{B(E2; 5/2_1^- \rightarrow 7/2_1^-)}{B(E2; 7/2_1^- \rightarrow 3/2_1^-)} = \frac{{( 5/2 K 2 0 | 7/2 K)}^2}{{( 7/2 K 2 0 | 3/2 K)}^2}.
\end{equation}
This ratio ($R$) of C.G. coefficients is defined only for $K=1/2$ and 3/2, yielding values of 0.1 and 1.0, respectively. The corresponding $B(E2)$ ratio is 0.4, which is closer to the $R$ value obtained for $K=1/2$, suggesting a $K=1/2^-$ rotational band. Under this approximation, the $0p0h$, $2p2h$, and $3p3h$ states are predicted to be of $K=3/2^+$, $K=1/2^+$, and $K=3/2^-$ bands, respectively. The rotational bands and their respective $K$ numbers, obtained with this simple argument, closely align with the experimental observations \cite{31Mg_ShapeCo2} and are consistent with the findings from angular momentum projected CCSD \cite{IOI_CC} and antisymmetrized molecular dynamics plus generator coordinate method (AMD+GCM) calculations \cite{31Mg_AMD}.

The quadrupole moments provide crucial insights into nuclear shape and deformation.  Considering $K$ as a good quantum number, the intrinsic quadrupole moments ($Q_0$) of a state with angular momentum $J$ can be written as \cite{KnumPRC2015, KnumNPA2018} 
\begin{equation}
    Q_s(J) = \frac{3K^2 - J(J+1)}{(J+1)(2J+3)} Q_0(J),
\end{equation}
where $Q_s$ denotes the spectroscopic quadrupole moment. A positive $Q_0$ ($Q_0>0$) indicates a prolate shape, whereas a negative value ($Q_0 <0$) corresponds to oblate deformation. The spectroscopic quadrupole moments of $3/2_1^+$, $3/2_1^-$, and $7/2_1^-$ states are 4.8 $e\mathrm{fm}^2$, -5.7 $e\mathrm{fm}^2$ and -12.9 $e\mathrm{fm}^2$, respectively, all indicating prolate deformation. The large quadrupole moment of the $7/2_1^-$ state suggests that it predominantly consists of strongly prolate-deformed configurations and co-exists with weakly deformed states at low excitation energies in $^{31}$Mg. In the experiments, the $K=1/2^+$ band belonging to $2p2h$ states appears as the ground state band, and the $K=3/2^+$ band of normal configurations is observed at an excited level. In the VS-IMSRG calculation, both bands are found to be altered. Notably, a similar prediction is also seen in the projected CCSD calculations \cite{IOI_CC}.

\textbf{$^{33}$Mg:} The band structures and configuration of low-lying excited states of $^{33}$Mg are displayed in Fig. \ref{fig:33MgE2mapAndStates}. Unlike $^{31}$Mg, the ground state band in $^{33}$Mg is well reproduced from VS-IMSRG calculations and predominantly of $2p2h$ character. The ground state band comprises of $3/2_1^-$, $7/2_1^-$, $5/2_1^-$, $1/2_1^-$, $9/2_1^-$ and $11/2_1^-$ states, and marked by strong $E2$ transitions as shown in left panel of Fig. \ref{fig:33MgE2mapAndStates}. A positive parity band, with its sequence of states $3/2_1^+$, $5/2_1^+$, $7/2_1^+$ and $9/2_1^+$, and interconnected with enhanced $E2$ strengths, is observed alongside the ground state band. 
Additionally, another positive parity band, comprising of $1/2_1^+$, $3/2_2^+$, and $5/2_2^+$ states and dominated by $3p3h$ configurations, is identified at a relatively higher excitation energy. Interestingly, no pure normal ($0p0h$) configurations states are found, yet certain states, such as $3/2_2^-$, $7/2_2^-$, and $3/2_3^-$, display nearly equal or a mixed composition of $0p0h$ and $2p2h$ configurations, with roughly $35-40\%$ $0p0h$ content. These states do not exhibit a clear band structure and primarily decay into the $3/2_1^-$ or $5/2_1^-$ states of the $2p2h$ band.

The $B(E2)$ ratios suggest a $K=3/2^-$ band for the $2p2h$ states, and the $1p1h$ and $3p3h$ states appear from $K=3/2^+$ and $K=1/2^+$ configurations, respectively. The IM-GCM calculations also predict $3/2^-$ as the ground state dominated by $K=3/2^-$ configuration. In $^{33}$Mg, the ground state is prolate deformed with a quadrupole moment $Q_s \sim 6.5$ $e\mathrm{fm}^2$. This quadrupole moment is found to be underestimated compared to the measured value 13.4(92) $e\mathrm{fm}^2$ \cite{33Mg_Qs}, possibly due to the missing higher-order collective excitations, as discussed in Ref. \cite{E2_strength_vsimsrg}. Above the ground state, the $7/2_1^-$, $5/2_1^-$, and $3/2_1^+$ states are almost degenerate with quadrupole moments -9.3 $e\mathrm{fm}^2$, -3.5 $e\mathrm{fm}^2$ and 7.7 $e\mathrm{fm}^2$, respectively. This indicates the presence of weak, moderate, and strong prolate-deformed shapes at low-excitation energies in $^{33}$Mg.


\section{Summary} \label{Sec4}
We have studied the $N=20$ island of inversion in the odd-mass Ne and Mg isotopes using the VS-IMSRG method, starting from chiral $NN$ and $NNN$ forces. The systematic trends of positive and negative parity yrast states towards the $N=20$ shell gap are well reproduced from VS-IMSRG calculations, indicating an inversion of the $p_{3/2}$ and $f_{7/2}$ orbitals compared to the conventional shell model and a reduced shell gap between $sd$ and $pf$-shell starting from $N=17$.
At $N=21$, where the shell gap disappears, the low-lying excited states are dominated by intruder $2p2h$ or $3p3h$ configurations. The calculated ground state magnetic moments closely align with the experimental data. The moments of the $3/2^-$ and $7/2^-$ states suggest that they appear from significant configuration mixing throughout the Ne and Mg isotopic chains. The rotational band structures are established in $^{31}$Mg and $^{33}$Mg based on $E2$ transition maps. In $^{31}$Mg, bands originating from both $1p1h$ and $2p2h$ configurations coexist with the $0p0h$ excitation ground state band. In contrast, the ground-state band of $^{33}$Mg is entirely of intruder nature, characterized by $2p2h$ excitations. Our results show good agreement with experimental data and are consistent with findings from other \textit{ab initio} methods. Furthermore, the quadrupole moments reveal the coexistence of diverse shapes, such as weak, moderate, and strongly prolate-deformed configurations at low excitation energies in $^{31}$Mg and $^{33}$Mg. The present work describes the island of inversion in odd-$A$ nuclei from first principles and offers valuable insights into the structure of neutron-rich Mg isotopes within this region.

\section*{Acknowledgment}
We sincerely thank Dr. T. Miyagi for availing the chiral $NN$ and $NNN$ matrix elements generated using NuHamil \cite{NuHamil} code. We also extend our gratitude to Dr. Y. Utsuno, Dr. T. Miyagi, and Dr. N. Shimizu for initial discussions and suggestions. S.S. would like to thank UGC (University Grant Commission), India, for providing financial support for his Ph.D. thesis work. P.C.S. acknowledges a research grant from SERB (India), CRG/2022/005167. We would like to thank the National Supercomputing Mission (NSM) for providing computing resources of `PARAM Ganga’ at the Indian Institute of Technology Roorkee, implemented by C-DAC and supported by the Ministry of Electronics and Information Technology (MeitY) and Department of Science and Technology (DST), Government of India.


\end{document}